\documentclass[aip,jcp,,preprint,letterpaper]{revtex4}
\usepackage{graphicx}
\usepackage{dcolumn}
\usepackage[caption=false]{subfig}
\usepackage{comment}
\usepackage{multirow}
\usepackage{array}
\usepackage{bm}
\usepackage{amsmath}
\usepackage{longtable}
\usepackage{color}
\usepackage{ textcomp }
\usepackage{diagbox}
\usepackage{chemformula}
\usepackage[version=3]{mhchem}
\usepackage{lipsum}

\begin{document}
\title{Graph Self-Supervised Learning for Optoelectronic Properties of Organic Semiconductors}

\author{Zaixi Zhang}
\affiliation{Anhui Province Key Lab of Big Data Analysis and Application, University of Science and Technology of China, Hefei, Anhui 230026, China}
\author{Qi Liu}
\email{qiliuql@ustc.edu.cn}
\affiliation{Anhui Province Key Lab of Big Data Analysis and Application, University of Science and Technology of China, Hefei, Anhui 230026, China}
\author{Shengyu Zhang} 
\affiliation{Tencent Quantum Lab, Shenzhen, Guangdong 518057, China}
\author{Chang Yu Hsieh}
\affiliation{Tencent Quantum Lab, Shenzhen, Guangdong 518057, China}
\author{Liang Shi}
\affiliation{Chemistry and Biochemistry, University of California, Merced, California 95343, United States}
\author{Chee Kong Lee}
\email{cheekonglee@tencent.com}
\affiliation{Tencent America, Palo Alto, CA 94306 , United States}

\begin{abstract}
The search for new high-performance organic semiconducting molecules is challenging due to the vastness of the chemical space, machine learning methods, particularly deep learning models like graph neural networks (GNNs), have shown promising potential to address such challenge. However, practical applications of GNNs for chemistry are often limited by the availability of labelled data. 
Meanwhile, unlabelled molecular data is abundant and could potentially be utilized to alleviate the scarcity issue of labelled data.
Here, we advocate the use of self-supervised learning to improve the performance of GNNs by pre-training them with unlabeled molecular data.
We investigate regression problems involving ground and excited state properties, both relevant for optoelectronic properties of organic semiconductors. Additionally, we extend the self-supervised learning strategy to molecules in non-equilibrium configurations which are important for studying the effects of disorder.
In all cases, we obtain considerable performance improvement over results without pre-training, in particular when labelled training data is limited, and such improvement is attributed to the capability of self-supervised learning in identifying structural similarity among unlabeled molecules. 
\end{abstract}
\maketitle

Organic semiconductors (OSCs) have been a vibrant field of research since the discovery of their electroluminescence properties in the 1960s and 1970s~\cite{Ostroverkhova2016, Kohler2015} due to their potential applications in solar cells~\cite{Hains2010,Myers2012,Lu2015,Hedley2017}, light-emitting devices\cite{Minaev2014,Xu2016} and field-effect transistors~\cite{Sirringhaus2014}.
The use of organic materials offers several advantages as compared to their inorganic counterparts, such as low production costs, versatile synthesis processes, and high portability. 
However the search of new high performance OSCs has proved challenging due to the vastness of chemical space.
Computational simulation could assist the search for OSCs materials with desirable electronic properties critical to their electronic applications at a lower cost compared to experiments.
Despite the efficiency of computational simulations, quantum chemistry methods such as the density functional theory (DFT) are still too expensive for high-throughput virtual screening involving a large number of candidate molecules~\cite{Hachmann2011}.
Recent successful applications of machine learning (ML) in chemistry show that it could accurately predict various molecular and material properties with vastly higher efficiency compared to quantum chemistry calculations~\cite{Smith2018, Christensen2021, Taylor2021, Kulichenko2021, Wang2021, Behler2011b, Behler2015, Behler2016,Dral2020,VonLilienfeld2020,Noe2020,  Olivares-Amaya2011,Sajeev2013a,Kanal2013,Shu2015,Li2015,Pyzer-Knapp2016,Pereira2017,Gomez-Bombarelli2016a,Musil2018,Mahapatra2018,Nagasawa2018,Jorgensen2018,Janai2018,Sahu2018,Padula2019,Padula2019b,Lee2019a,St.John2019,Atahan-Evrenk2019,Bian2019,Lederer2019,Roch2020,Simine2020,Lu2020,farahvash2020machine, Duan2020, Welborn2018, Prezhdo2020, Poltavsky2021}. 
In particular, state-of-the-art deep learning methods such as the graph neural networks (GNNs) have shown to be able to achieve prediction accuracy superior to other traditional ML methods~\cite{Ramakrishnan2014, Duvenaud2015, Schutt2017, Gilmer2017, Wu2018, Lu2019a, Schutt2018, Schutt2019, Chen2019, Unke2019, Klicpera2020, Liu2020, Qiao2020, hao2020asgn}. 

Despite its immense potential, practical application of GNNs in chemistry is frequently limited by the availability of labelled training data. Meanwhile, in many cases unlabelled data are abundant, e.g. from publicly available database like PubChem or molecular dynamics simulations. In order to utilize the availability of these unlabelled data and overcome the scarcity of labelled data, recently various self-supervised pre-training strategies have been devised for GNNs, and have been successfully demonstrated in social network and biological domains~\cite{Xie2021, Lu2021, Hu*2020Strategies, Rong2020, Zhang2021}. However its applications on quantum mechanical properties have been limited and only available on simple small molecules like those in the QM7 and QM8 datasets~\cite{Rong2020}. Additionally, these pre-training strategies have not been tested on excited state properties or molecules in non-equilibrium configurations.
In this work we advocate the use of self-supervised learning (SSL) in GNNs for predicting the optoelectronic properties of OSCs. 
SSL pre-training of GNNs consists of two steps: unsupervised learning and supervised fine-tuning. During the unsupervised learning stage, a GNN is first trained on a large collection of unlabeled molecular data such that it derives generic transferable knowledge encoding the intrinsic graph representation of molecules. During the fine-tuning stage the pre-trained GNN model is fine-tuned on task-specific molecular data, such that it adapts the generic knowledge for specific tasks.

For the first application in this work, we apply SSL to the prediction of optoelectronic properties of organic photovoltaic molecules where the only input is the SMILES strings of the molecules.
For the second application, we extend the SSL strategy to molecules in non-equilibrium configurations by incorporating 3D coordinates into the training of SSL. Existing SSL studies only focus on molecules in their equilibrium geometries and thus do not consider the effect of disorder or temperature on the electronic properties of OSC molecules. However the presence of disorder could have significant implication on the performance of OSC devices. For example it is known that the existence of disorder can limit the transport of charges and excitons~\cite{Lee2019b}, leading to a drop in device efficiency. On the other hand disorder can sometimes assist the separation of charge-transfer exciton, an important step for efficient organic photovoltaics~\cite{Deotare2015, Shi2017}. 
Thus to design high-performance OSCs, it is essential to understand the effects of disorder on the electronic properties of OSCs. Therefore in this work, we also explore the application of SSL on predicting the excited state properties of OSC molecules in non-equilibrium configurations.

\begin{figure*}[ht!]
  \includegraphics[width=5.5in]{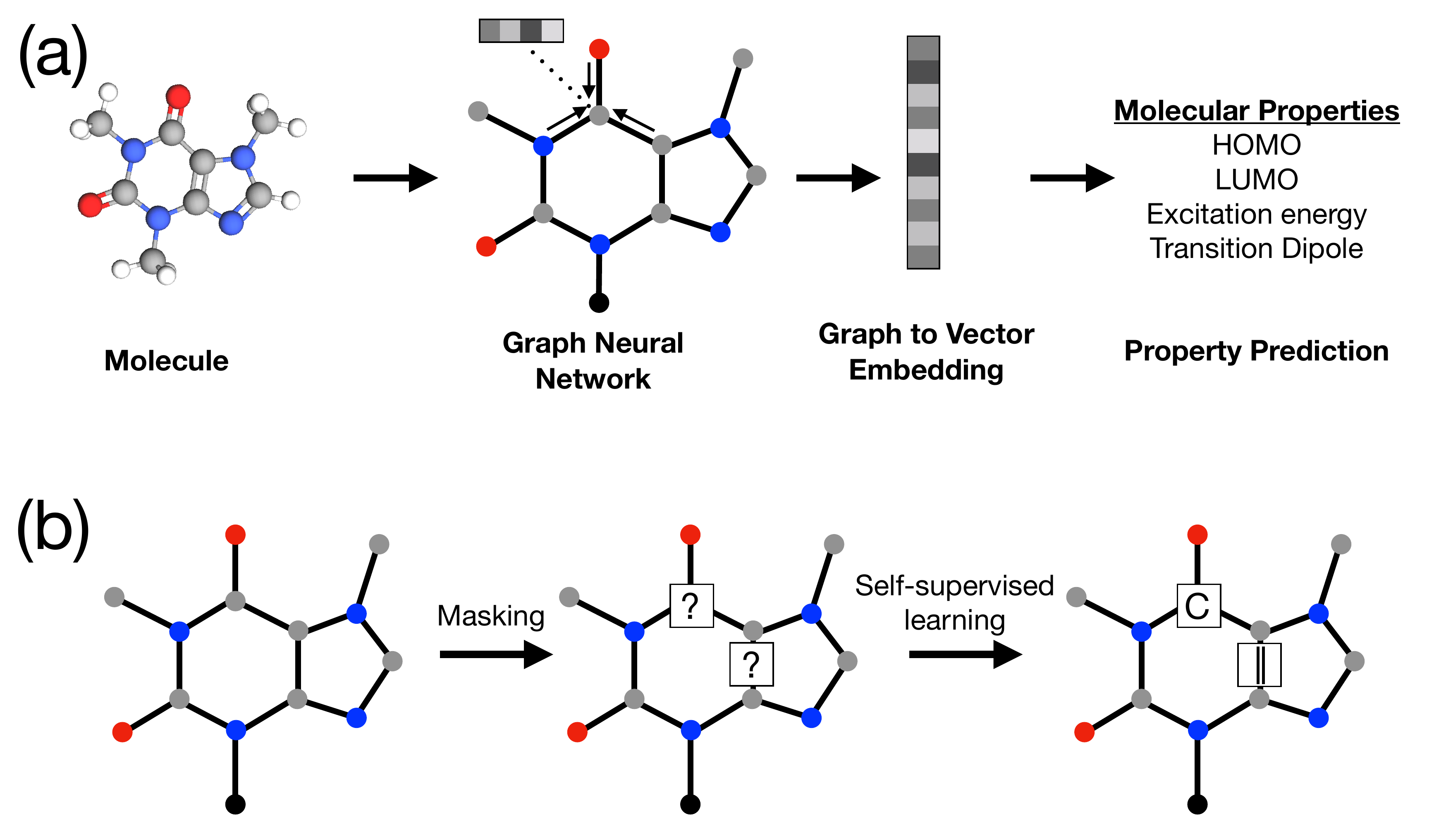}
    \caption{A schematic of the model used in this work. (a) A molecule is first converted into a computational graphs, and each node (i.e. atom) is represented by an embedding vector. 
    The optimal node embedding is learned via a message passing algorithm, i.e. the embedding vector is iteratively updated by aggregating the embeddings of its neighboring nodes and edges. 
    The molecule-level embedding vectors can then be generated by pooling all the atoms through summation. 
    The learned molecular representation can be used for the prediction of molecular properties through the read-out phase.
    (b) In self-supervised learning (SSL), we randomly mask 15\% of the node (i.e atom) and edge (i.e. bond) attributes, and the GNNs are tasked to predict these masked attributes.}
    \label{fig:schematic}
\end{figure*}

\textit{Graph neural networks} - 
In contrast with traditional ML methods where hand-crafted molecular descriptors are required as input, deep learning methods such as GNNs are capable of extracting informative  representation of a molecule solely from atom types and Cartesian coordinates.
In GNNs, the basic chemical information of molecules are encoded as computational graphs and these graphs are used as the input for the graph-based training algorithm. As compare to the traditional ML methods, GNNs are capable of representing the irregular molecular graph structures more naturally. Specifically, a computational graph $G = (V, E)$ is defined as the connectivity relations between a set of nodes ($V$) and a set of edges ($E$). Naturally, a molecule can also be considered as a graph consisting of a set of atoms (nodes) and a set of bonds (edges).

A schematic of the GNNs is shown in Fig.~\ref{fig:schematic}a. After a molecule is converted into a computational graph, each node (atom) is represented by an embedding vector. 
GNNs learn the optimal representation of each atom  using a message passing algorithm that iteratively aggregates the information of its neighboring atoms and the corresponding edges~\cite{Gilmer2017}. 
After the message passing phase, the molecule-level embedding vectors can be generated by pooling all the atoms through summation.
Finally, the learned molecular representation can be used for the prediction of molecular properties through the read-out phase. 

\textit{Self-supervised learning} - 
We use node and edge-level attribute masking for SSL of GNNs in this work: first the node and edge attributes of the graphs are masked, then the GNNs are tasked to predict those attributes based on neighboring structures~\cite{Hu*2020Strategies}. 
Fig.\ref{fig:schematic} (b) illustrates the working mechanism of attribute masking when applied to a molecular graph. We randomly mask the atom and bond types of the molecular graphs by replacing them with special masked indicators. GNNs are then tasked to predict these masked node or edge at attributes. More details of the attribute masking SSL can be found in the Supplementary Materials (SM). After the GNN pre-training is finished, we then fine-tune the pre-trained GNN model on specific prediction tasks.

\begin{figure*}[ht!]
  \includegraphics[width=6.5in]{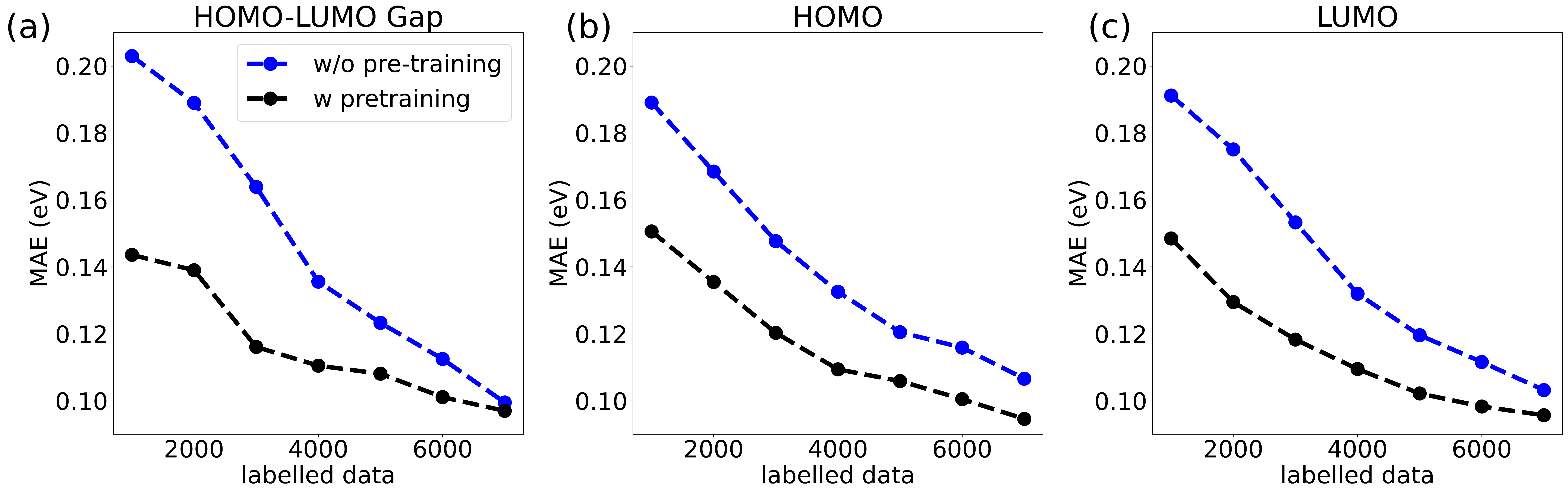}
    \caption{Test set mean absolute errors (MAE) of (a) HOMO-LUMO gap, (b) HOMO and (c) LUMO of organic photovoltaics (OPV) molecules as a function of the number of labelled training data. Blue lines represent the results from direct training of GNNs without  pre-learning, whereas the black lines denote the results of GNNs pre-train with unlabelled data.}
    \label{fig:opv}
\end{figure*}

\textit{Organic Photovoltaic Dataset} - 
We first apply the attribute masking SSL strategy to predict the quantum property of organic photovoltaic (OPV) molecules.
The OPV dataset used in this work contains the SMILES strings of the 90823 unique molecules and their corresponding the ground state electronic properties obtained from DFT calculations with B3LYP/6-31G(d)~\cite{St.John2019}. 5000 molecules were randomly selected from the dataset for each of the validation and test sets. The remaining data is used for pre-training and fine-tuning. 
The underlying GNN used is the Graph Isomorphism Network~\cite{Xu2019}, a powerful GNN that is widely used in a variety of graph related task. However it is worth noting that the pre-training strategy advocated in this work is general and applicable to most GNNs. 
The entire training process consists of pre-training and fine-tuning. We first pre-train the GNNs with the entire training dataset (without the molecules in the test and validation sets) using the attribute masking SSL strategy as depicted in Fig.~\ref{fig:schematic} (b). 
After pre-training, we then fine-tuned the GNNs with just a small number of the labelled data. The details of the training process and parameters can be found in the SM.

The ground state electronic properties we focus on are the values of HOMO-LUMO gap, HOMO and LUMO, the corresponding results are shown in Fig.~\ref{fig:opv} in which the test set mean absolute errors (MAEs) of the property predictions (in eV) are shown as a function of the number of labelled training data used in the fine-tuning stage. We use all 80823 unlabelled molecules for the pre-training. 
The prediction results with and without SSL pre-training are shown as black and blue dots with dashed lines, respectively. 
It can be seen that the use of SSL pre-training generally leads to considerable improvement in prediction accuracy.
For example with 1000 labelled training data, the MAE for HOMO-LUMO gap prediction drops from $0.203$eV to $0.144$eV, an approximately 30\% improvement in accuracy. From another point of view, GNN without training would need approximately four times the number of labelled training data to achieve the same performance as the GNN with pre-training. 
Sizable but smaller relative improvements (approximately 20\% reduction in MAEs with 1000 labelled data) can also be observed for the predictions of HOMO and LUMO values. 
As expected, it can also be seen that the relative improvement decreases when the number of labelled data increases for all three properties, which shows that SSL is most useful for applications when labelled data is scarce. 

\begin{figure}[ht!]
  \includegraphics[width=2.5in]{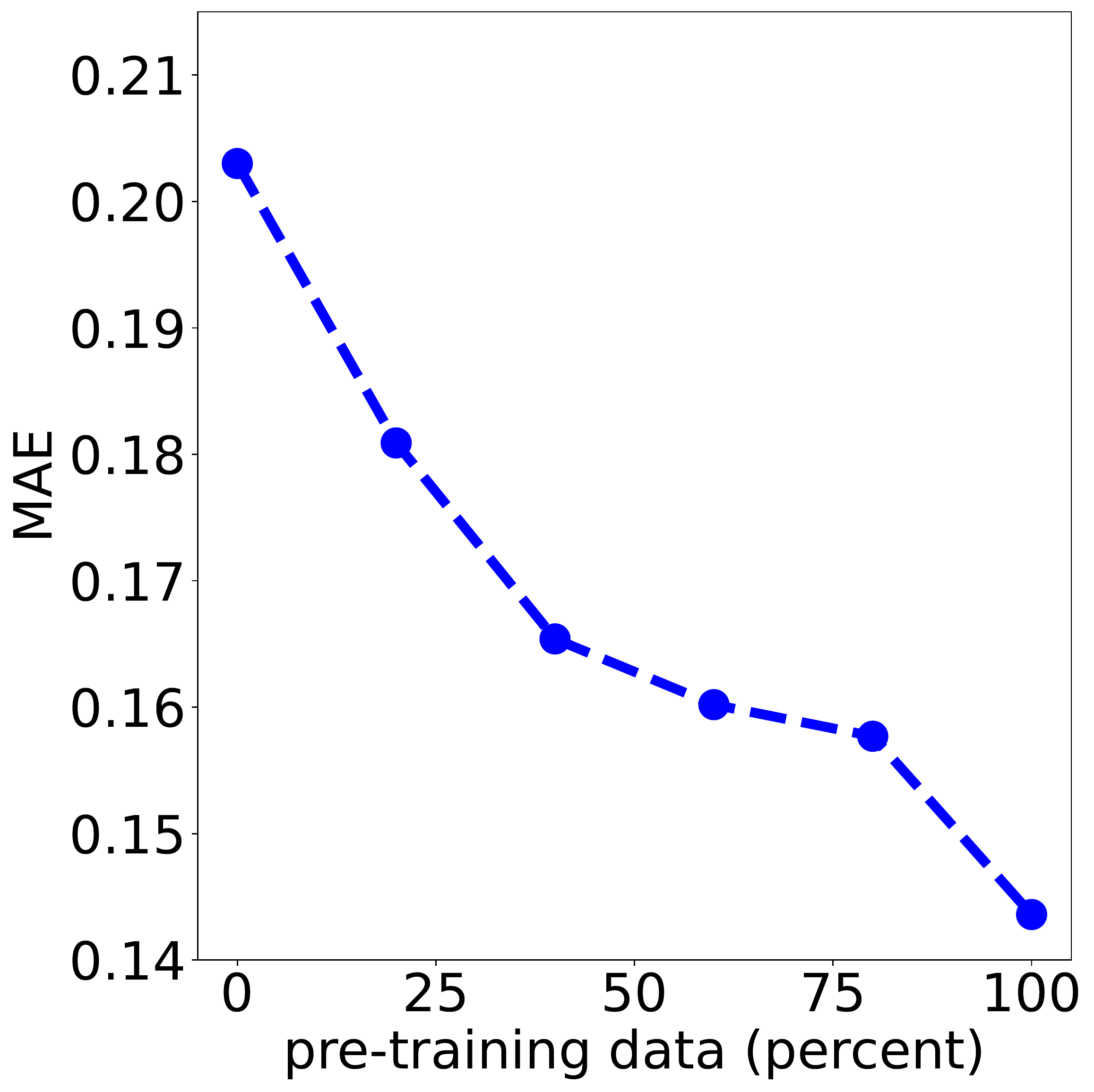}
    \caption{Test set MAE of HOMO-LUMO gap for OPV dataset as a function of the number of unlabelled pre-training data. There are a total of 80823 molecules in the pre-training dataset.}
    \label{fig:data}
\end{figure} 

Next we investigate the performance dependence of SSL on the amount of unlabelled data used in the pre-training stage. Fig.~\ref{fig:data} shows the test set MAE of HOMO-LUMO gap prediction as a function of the number of unlabelled pre-training data, expressed in terms of the percentage of the total pre-training dataset (80823 molecules). We use 1000 labelled data for the fine-tuning. 
As expected, the prediction performance is sensitive to the amount of data used in the pre-training of GNNs, more data leads to higher performance. 
Interestingly, it is shown that the MAE curve has not leveled even after using all of the training data for pre-training, indicating the performance of SSL can be further improved from more unlabelled data. 
One could potentially pre-train GNNs using other large publicly available chemical datasets such the PubChem~\cite{Kim2021} and ZINC~\cite{Irwin2012} datasets in addition to the OPV dataset, each of these datasets contains millions of molecules and could potentially further boost the performance of the GNNs. 
Though pre-training of GNNs with such a large dataset could be computationally intensive, it needs only to be performed once and the pre-trained GNNs can be used for various downstream tasks. Gigantic pre-trained models have been released in the computer vision~\cite{Simonyan2015, Szegedy2015, AlexNet2012} and nature language processing~\cite{Devlin2019, Alex2019} domains, and had subsequently led to rapid advances in these fields, it will be a fruitful endeavor to attempt similar approach in chemical sciences in the future.

\begin{figure}[ht!]
  \includegraphics[width=3.5in]{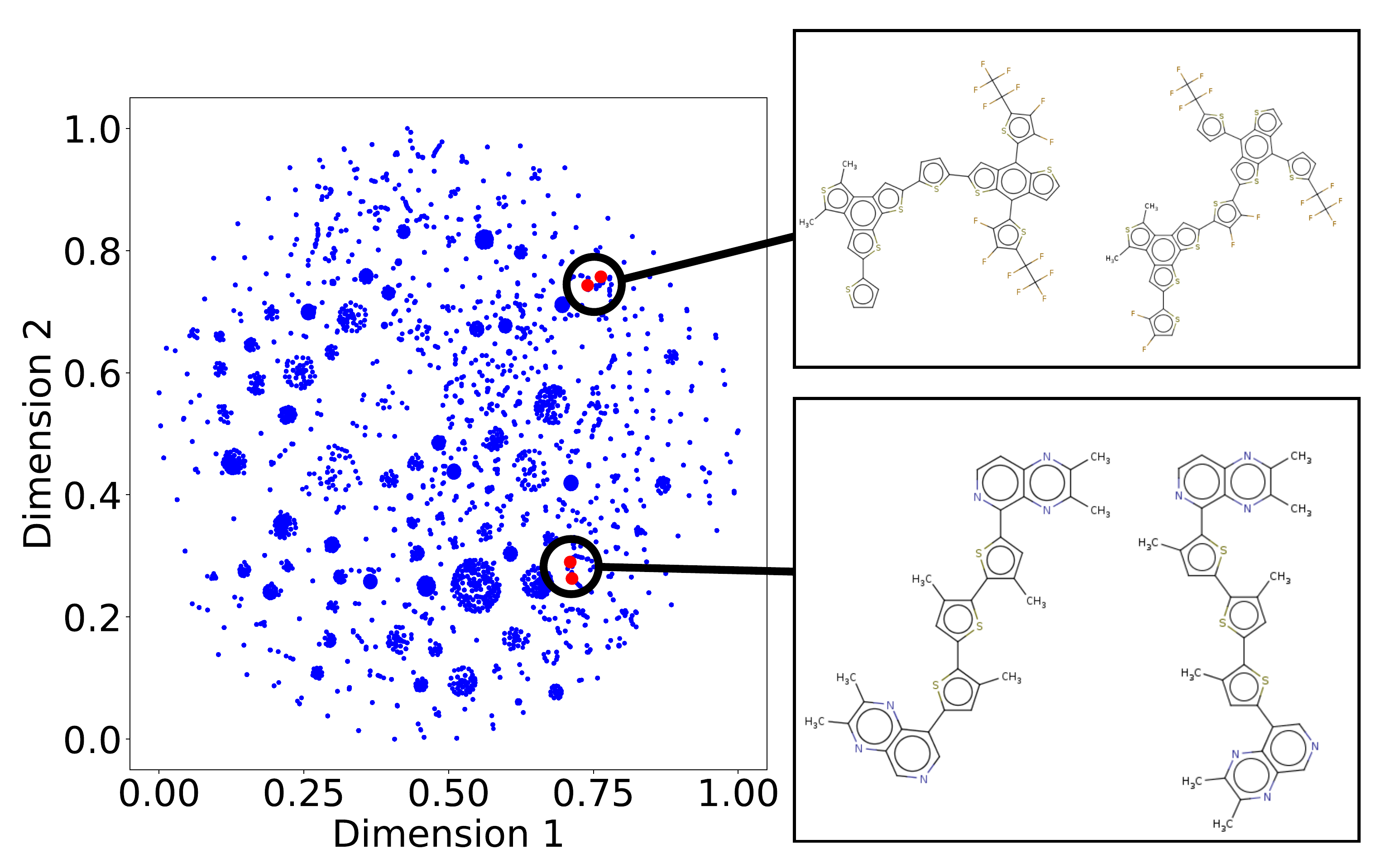}
    \caption{Two-dimensional visualization of the results of the pre-training using t-SNE.}
    \label{fig:embedding}
\end{figure}

\textit{Embedding Visualization} - 
To better understand the effect of pre-training, we use the t-distributed Stochastic Neighbor Embedding (t-SNE) technique to visualize the GNN vector embeddings after SSL. t-SNE is an unsupervised dimensionality reduction technique commonly used for the visualization of high-dimensional datasets. As a non-linear dimensionality technique, t-SNE reduces the dimensions of correlated data by projecting the original set of vectors onto small number of principal components while preserving most of the data variation. 
Fig.~\ref{fig:embedding} shows the two-dimensional distribution of the molecule vector embeddings after performing t-SNE. 
It can be seen that the embeddings of many molecules form clusters in the t-SNE distribution. It is expected that molecules in a cluster or nearby molecules in the reduced dimensions share some structural similarity.
By looking into the structures of some of the adjacent molecules, we indeed find some nearby molecules that are structurally very similar, as illustrated in Fig.~\ref{fig:embedding}. 
However due to the diversity and complexity of the OPV dataset, many molecules are isolated and do not belong to any cluster in the reduced dimensions.

\textit{Non-equilibrium configurations} - 
We next explore the capability of SSL on molecules in non-equilibrium configurations and their excited state properties. 
For this purpose we use a dataset that contains 80000 non-equilibrium configurations of sexithiophene molecule. These configurations are generated from molecular dynamics simulations at 1000K and the excited properties are obtained using time-dependent DFT (TD-DFT) calculations with the CAM-B3LYP functional, more details about the dataset can be found in Refs.~\citenum{Lu2020} and \citenum{Lee2021}.
Similar to the OPV calculations, we randomly select 5000 configurations each for validation and testing, and use the remaining data for pre-training and fine-tuning. 
The underlying GNN used is SchNet since we need to take the 3D molecular coordinates into account in addition to the graph structures~\cite{Schutt2018}. 
Since the non-equilibrium dataset involves the same molecule, the standard atom/bond type masking in SSL is not applicable. Instead, we mask the inter-atomic distance feature vectors after the rbf layer in the filter-generating network, and SchNet is tasked to predict these pair-wise vector in the pre-training stage.

\begin{figure}[ht!]
  \includegraphics[width=2.5in]{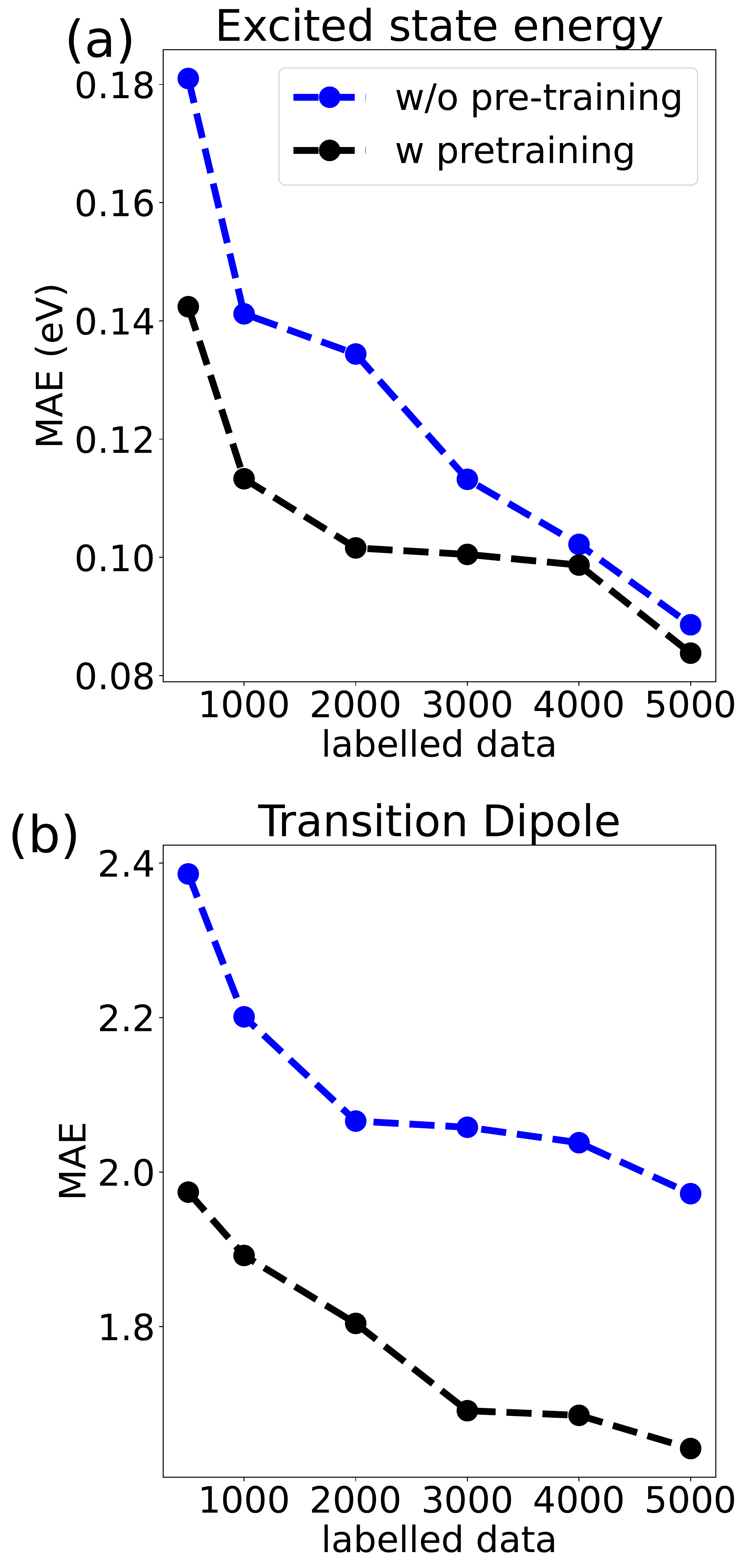}
    \caption{Evaluation on excited state properties of non-equilibrium configurations. Test set MAEs of (a) excited state energy and (b) magnitude of transition dipole moment of sexithiophene molecule as a function of the number of labelled training data. Blue lines represent the results from direct training of SchNet without pre-learning, whereas the black lines denote the results of SchNet pre-trained with unlabelled data.}
    \label{fig:T6}
\end{figure}

In Fig.~\ref{fig:T6} we evaluate the performance of pre-training in predicting two excited state properties: the lowest excited state energy and the magnitude of its transition dipole moment. 
It can be seen from Fig.~\ref{fig:T6}a that the prediction of the excited state energy is considerably improved by the use of SSL pre-training as compared to the results without pre-training. Similar to the results with OPV dataset, the improvement is most significant when the labelled data is scarce. For example, with only 1000 labelled data, the MAEs drops from $1.81$eV to $1.42$eV, an approximately $22\%$ reduction. As the number of labelled training increases, the improvement decreases and becomes nearly negligible beyond 4000 labelled data. 
Next in Fig.~\ref{fig:T6}b, we show the test set MAE of transition dipole moment magnitude prediction as a function of the number of labelled training data. We again observe significant improvement from SSL pre-training of GNNs.
Interestingly, unlike the excited state energy prediction, the magnitude of improvement of nearly $16 - 20\%$ is nearly constant even when the labelled training data increases from 1000 to 5000. It has been previously shown that the prediction of transition dipole moment is more difficult compared to other electronic properties~\cite{Lu2020, Ye2019}, our results here suggest SSL could be especially useful for the prediction of such challenging property in which the need for training data is greater.

\textit{Discussions and Conclusions} - 
We demonstrate the capability of SSL pre-training of GNNs in improving the prediction accuracy of optoelectronic properties of OSCs.
Similar to pre-training strategies in other domains such as computer vision where the neural networks are tasked to learn the basic features such as edges and curves in pictures from abundant unlabelled data, the SSL pre-training allows GNNs to recognize the basic structures in molecules, such as bonds and ring structures. From the t-SNE plot in Fig.~\ref{fig:embedding}, it can be seen that some structurally similar molecules are grouped together in the vector embedding space during the process of pre-training, such grouping assist the training process during the fine-tuning stage.  Importantly, the pre-training only needs to be performed once, and the pre-trained GNNs can be fine-tuned for any property-specific task. 

In this work we apply SSL to two types of problems, namely equilibrium ground state and non-equilibrium excited state property predictions.
For the equilibrium case, the only input required is the SMILES strings of the molecules. A potential application is the virtual screening of OSC molecules for desirable optical properties before they are actually synthesized. 
For the non-equilibrium counterpart, accurate prediction of the dependence of the OSC optoelectronic properties on the 3D configuration is crucial in understanding how disorder could affect the performance of OSCs. 
In both cases, we obtain considerable performance improvement over results without the use of pre-training, and the improvement is most significant when labelled training is scarce.
Finally, there are other ML strategies that could alleviate the need of labelled data, e.g. transfer learning and active learning, and it will be a fruitful endeavor to combine SSL with these strategies to maximize the potential of ML methods for chemical applications.

\bibliographystyle{naturemag_noURL.bst}
\bibliography{MyCollection}

\end{document}


\newcommand{\wn}{cm$^{-1}$}
\newcommand{\td}{$\sim$}
\newcommand{\la}{\langle}
\newcommand{\ra}{\rangle}
\newcommand{\e}{\epsilon}
\newcommand{\w}{\omega}
\newcommand{\bracket}[1]{\left\langle #1 \right\rangle}
\newcommand{\degreec}{^{\circ}{\rm C}}
\newcommand{\be}{\begin{equation}}
\newcommand{\ee}{\end{equation}}
\newcommand{\ie}{{\it i.e.}}
\newcommand{\eg}{{\it e.g.}}
\newcommand{\etal}{{\it et al.}}
\newcommand{\bra}[1]{\left<#1\right|}
\newcommand{\ket}[1]{\left|#1\right>}
\newcommand{\ketbra}[2]{\ket{#1}\bra{#2}}

\title{{\huge Supporting Materials}\\
\vskip 0.2in
Graph Self-Supervised Learning for Optoelectronic Properties of Organic Semiconductors}

\author{Zaixi Zhang}
\affiliation{Anhui Province Key Lab of Big Data Analysis and Application, University of Science and Technology of China, Hefei, Anhui 230026, China}
\author{Qi Liu}
\email{qiliuql@ustc.edu.cn}
\affiliation{Anhui Province Key Lab of Big Data Analysis and Application, University of Science and Technology of China, Hefei, Anhui 230026, China}
\author{Shengyu Zhang} 
\affiliation{Tencent Quantum Lab, Shenzhen, Guangdong 518057, China}
\author{Chang Yu Hsieh}
\affiliation{Tencent Quantum Lab, Shenzhen, Guangdong 518057, China}
\author{Liang Shi}
\affiliation{Chemistry and Biochemistry, University of California, Merced, California 95343, United States}
\author{Chee Kong Lee}
\email{cheekonglee@tencent.com}
\affiliation{Tencent America, Palo Alto, CA 94306 , United States}
\maketitle
\section{Implementation Details of Graph Neural Networks}
\subsection{Implementation of GIN}
We apply Graph Isomorphism Network (GIN) \cite{Xu2019} for the molecular property prediction of OPV dataset.
We select the following settings for GIN: 300 dimensional hidden units, 5 GNN layers, ReLU activation, dropout probability 0.5 for all layers except the input layer, and average
pooling for the readout function. 

For the input to GIN, we use RDKit \cite{landrum2006rdkit} to obtain the node and edge features: Node features: Atomic number; Chirality tag: $\{$unspecified, tetrahedral cw, tetrahedral ccw, other$\}$. Edge features: Bond type: $\{$single, double, triple, aromatic$\}$; Bond direction: $\{$–, endupright, enddownright$\}$. 

GIN is trained with Adam optimizer with a learning rate of 0.001. Both pretraining and finetuning are trained for 100 epochs. For self-supervised pre-training, we use a batch size of 256, while for finetuning, we use a batch size of 32. We use Pytorch and Pytorch Geometric for all of our implementation. 
\subsection{Implementation of Schnet}
We apply Schnet \cite{Schuett2017a} for the property prediction of non-equilibrium thiophene molecules. We select the following settings for Schnet: 2 interaction blocks, 64 dimensional hidden units, shifted softplus activation, 3.0 cutoff, 0.1 width, and sum pooling for the readout function.
The input to Schnet are the atom attributes, i.e. atomic numbers and the interatomic distances.

Schnet is trained with Adam optimizer with a learning rate of 0.0001. We pretrain Schnet for 100 epochs and finetune it for 1000 epochs. For both self-supervised pre-training and finetuning, we use a batch size of 20. We use Pytorch and Pytorch Geometric for all of our implementation. 

\section{Implementation Details of Self-supervised Learning}
We employ attribute masking as the strategy for self-supervised learning~\cite{Hu*2020Strategies}. Generally, we mask the node/edge attributes and let GNNs predict those attributes based on neighboring structure. In experiments, we randomly sample 15$\%$ nodes and edges and replace them with special mask indicators. For SchNet, we replace the interatomic distance embeddings of m
asked edges after the rbf layer with special mask indicators. We then apply GNNs to obtain
the corresponding node/edge embeddings (edge embeddings can be obtained as a sum of node embeddings of the edge’s end nodes). Finally, a fully-connected layer is applied on top of embeddings to predict the masked node/edge attributes. For the masked edges in Schnet, we predict which interval the interatomic distance belongs to.

\section{Dependence on Pre-training Data Size}
\begin{figure}[ht!]
  \includegraphics[width=4.5in]{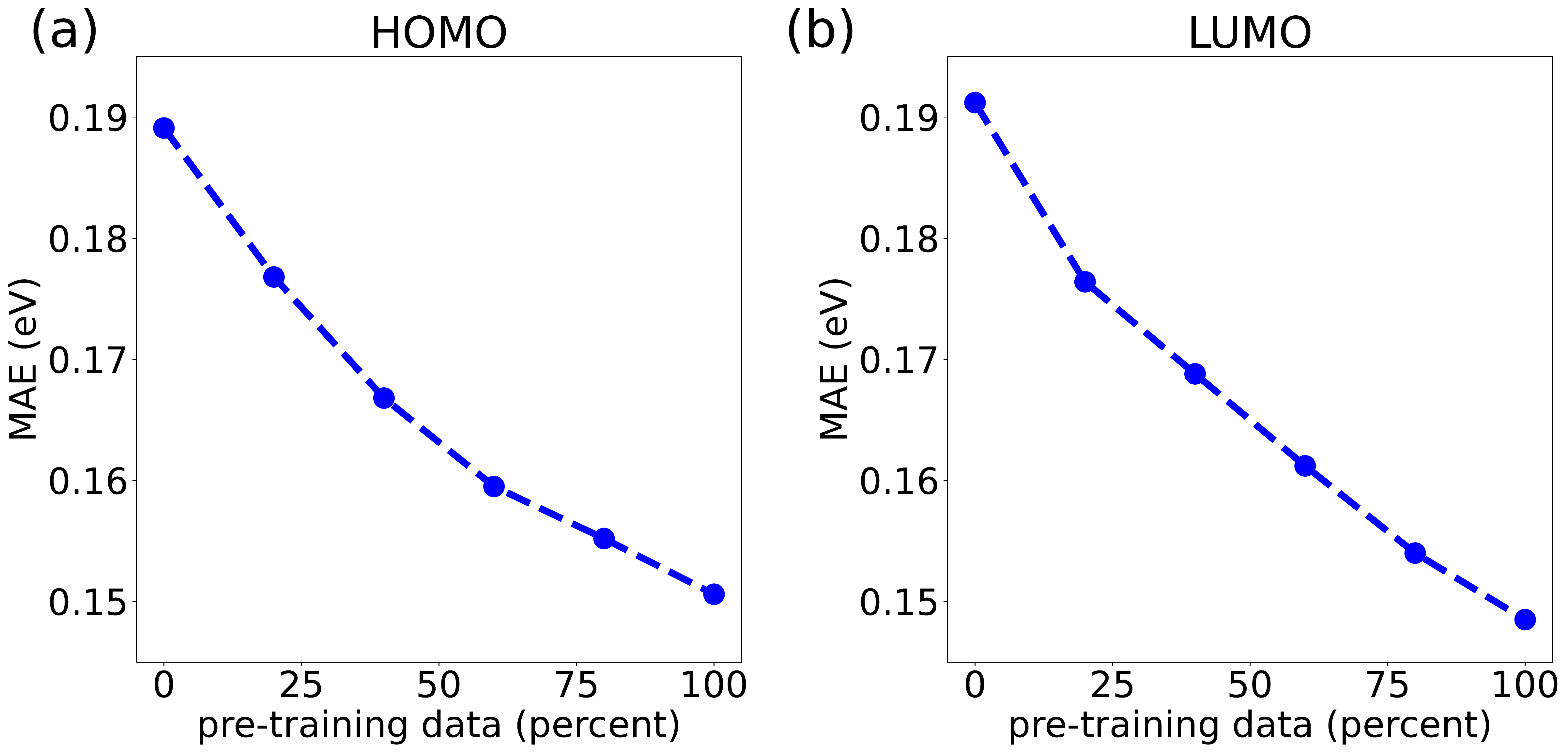}
    \caption{Test set MAEs of HOMO and LUMO for OPV dataset as a function of the number of unlabelled pre-training data. There are a total of 80823 molecules in the pre-training dataset.}
    \label{fig:data_homo_lumo}
\end{figure}

In Fig.\ref{fig:data_homo_lumo}, we show additional results of  the  performance  dependence  of SSL  on  the  amount  of  unlabelled  data  used  in  the pre-training  stage. Similar to  Fig.~3 in the main text, 1000 labelled data is used for the fine-tuning stage. 

\clearpage
\bibliographystyle{naturemag_noURL}
\bibliography{MyCollection}